\def\fileversion{v2.0}
\def\filedate{5 May 1992}

\newdimen\@bls                    
\@bls=\baselineskip               
\advance\@bls -1ex                
\newdimen\@eps                    %
\def\section{\@startsection{section}{1}{\z@}
  {2\@bls plus 0.5\@bls}{1\@bls}{\normalsize\bf}}
\def\subsection{\@startsection{subsection}{2}{\z@}
  {1\@bls plus 0.25\@bls}{\@eps}{\normalsize\bf}}
\def\subsubsection{\@startsection{subsubsection}{3}{\z@}
  {1\@bls plus 0.25\@bls}{\@eps}{\normalsize\bf}}
\def\paragraph{\@startsection{paragraph}{4}{\parindent}
  {1\@bls plus 0.25\@bls}{0.5em}{\normalsize\bf}}
\def\subparagraph{\@startsection{subparagraph}{4}{\parindent}
  {1\@bls plus 0.25\@bls}{0.5em}{\normalsize\bf}}

\def\@sect#1#2#3#4#5#6[#7]#8{\ifnum #2>\c@secnumdepth
  \def\@svsec{}\else
  \refstepcounter{#1}\edef\@svsec{\csname the#1\endcsname.\hskip0.5em}\fi
  \@tempskipa #5\relax
  \ifdim \@tempskipa>\z@
    \begingroup
      #6\relax
      \@hangfrom{\hskip #3\relax\@svsec}{\interlinepenalty \@M #8\par}%
    \endgroup
    \csname #1mark\endcsname{#7}\addcontentsline
      {toc}{#1}{\ifnum #2>\c@secnumdepth \else
        \protect\numberline{\csname the#1\endcsname}\fi #7}%
  \else
    \def\@svsechd{#6\hskip #3\@svsec #8\csname #1mark\endcsname
      {#7}\addcontentsline{toc}{#1}{\ifnum #2>\c@secnumdepth \else
        \protect\numberline{\csname the#1\endcsname}\fi #7}}%
  \fi \@xsect{#5}}

\long\def\@makefigurecaption#1#2{\vskip 10mm #1. #2\par}

\long\def\@maketablecaption#1#2{\hbox to \hsize{\parbox[t]{\hsize}
  {#1 \\ #2}}\vskip 0.3ex}

\def\fnum@figure{Figure \thefigure}
\def\figure{\let\@makecaption\@makefigurecaption \@float{figure}}
\@namedef{figure*}{\let\@makecaption\@makefigurecaption \@dblfloat{figure}}

\def\table{\let\@makecaption\@maketablecaption \@float{table}}
\@namedef{table*}{\let\@makecaption\@maketablecaption \@dblfloat{table}}

\textfloatsep=\floatsep           
\intextsep=\floatsep              

\long\def\@makefntext#1{\parindent 1em\noindent\hbox{${}^{\@thefnmark}$}#1}

\def\maketitle{\begingroup        
    \def\thefootnote{\fnsymbol{footnote}}%
    \newpage \global\@topnum\z@
    \@maketitle \thispagestyle{plain}\@thanks
  \endgroup
  \let\maketitle\relax \let\@maketitle\relax
  \gdef\@thanks{}\let\thanks\relax
  \gdef\@address{}\gdef\@author{}\gdef\@title{}\let\address\relax}

\def\justify@on{\let\\=\@normalcr
  \leftskip\z@ \@rightskip\z@ \rightskip\@rightskip}

\newbox\fm@box                    

\def\@maketitle{
  \global\setbox\fm@box=\vbox\bgroup
    \raggedright                  
    \hyphenpenalty\@M             
    {\Large \@title \par}         
    \vskip\@bls                   
    {\normalsize                  
     \@author \par}               
    \vskip\@bls                   
    \@address                     
  \egroup
  \twocolumn[
    \unvbox\fm@box                
    \vskip\@bls                   
    \unvbox\abstract@box          
    \vskip 2pc]}                  

\newcounter{address}
\def\theaddress{\alph{address}}
\def\@makeadmark#1{\hbox{$^{\rm #1}$}}

\def\address#1{\addressmark\begingroup
  \xdef\@tempa{\theaddress} \let\\=\relax
  \def\protect{\noexpand\protect\noexpand}\xdef\@address{\@address
  \protect\addresstext{\@tempa}{#1}}\endgroup}
\def\@address{}

\def\addressmark{\stepcounter{address}%
  \xdef\@tempa{\theaddress}\@makeadmark{\@tempa}}

\def\addresstext#1#2{\leavevmode \begingroup
  \raggedright \hyphenpenalty\@M \@makeadmark{#1}#2\par \endgroup
  \vskip\@bls}

\newbox\abstract@box              

\def\abstract{%
  \global\setbox\abstract@box=\vbox\bgroup
  \small\rm
  \ignorespaces}
\def\endabstract{\par \egroup}

\def\thebibliography#1{\section*{REFERENCES}\list{\arabic{enumi}}
  {\settowidth\labelwidth{#1}\leftmargin=1.67em
   \labelsep\leftmargin \advance\labelsep-\labelwidth
   \itemsep\z@ \parsep\z@
   \usecounter{enumi}}\def\makelabel##1{\rlap{##1}\hss}%
   \def\newblock{\hskip 0.11em plus 0.33em minus -0.07em}
   \sloppy \clubpenalty=4000 \widowpenalty=4000 \sfcode`\.=1000\relax}

\newcount\@tempcntc
\def\@citex[#1]#2{\if@filesw\immediate\write\@auxout{\string\citation{#2}}\fi
  \@tempcnta\z@\@tempcntb\m@ne\def\@citea{}\@cite{\@for\@citeb:=#2\do
    {\@ifundefined
       {b@\@citeb}{\@citeo\@tempcntb\m@ne\@citea
        \def\@citea{,\penalty\@m\ }{\bf ?}\@warning
       {Citation `\@citeb' on page \thepage \space undefined}}%
    {\setbox\z@\hbox{\global\@tempcntc0\csname b@\@citeb\endcsname\relax}%
     \ifnum\@tempcntc=\z@ \@citeo\@tempcntb\m@ne
       \@citea\def\@citea{,\penalty\@m\ }
       \hbox{\csname b@\@citeb\endcsname}%
     \else
      \advance\@tempcntb\@ne
      \ifnum\@tempcntb=\@tempcntc
      \else\advance\@tempcntb\m@ne\@citeo
      \@tempcnta\@tempcntc\@tempcntb\@tempcntc\fi\fi}}\@citeo}{#1}}

\def\@citeo{\ifnum\@tempcnta>\@tempcntb\else\@citea
  \def\@citea{,\penalty\@m\ }%
  \ifnum\@tempcnta=\@tempcntb\the\@tempcnta\else
   {\advance\@tempcnta\@ne\ifnum\@tempcnta=\@tempcntb \else \def\@citea{--}\fi
    \advance\@tempcnta\m@ne\the\@tempcnta\@citea\the\@tempcntb}\fi\fi}

\documentstyle[twoside,fleqn,proc]{article}

\def \fhat{\hat{F}}

\newcommand{\AmS}{{\protect\the\textfont2
  A\kern-.1667em\lower.5ex\hbox{M}\kern-.125emS}}

\title{Results in the static approximation}

\author{C. Alexandrou\address{Paul Scherrer Institute, CH-5232 Villigen,
        Switzerland}
        in collaboration  with\\
        S. G\"usken\address{Physics Department, University of Wuppertal,
           D-5600 Wuppertal 1,
         Fed. Rep. Germany},
        F. Jegerlehner$^{\rm a}$, K. Schilling$^{\rm b}$ and
        R. Sommer\address{CERN, Theory Division, CH-1211 Geneva 23,
       Switzerland}
       \\}

\begin{document}

\begin{abstract}
We present a comprehensive study of finite volume and lattice spacing effects
on observables calculated in the static approximation.
Ground state projection using smearing techniques is studied at $\beta=6.26$.
We give high statics results for
the pseudoscalar decay constant after extrapolating
to the chiral and continuum limit. In addition results for mass splittings and
the
distance of string breaking in the $Q\bar{Q}$ potential are presented.
\end{abstract}

\maketitle

\section{Introduction}

For the B system it seems very natural to apply the infinite mass effective
theory (IMET) on the lattice \cite{Eich} to calculate interesting quantities
like the pseudoscalar decay
constant $f$ which has not been measured experimentally so far. A related
question which can be studied on the lattice is beyond which mass IMET
can be considered a reasonable approximation.
Since IMET has important  phenomenological implications studying the static
limit on the lattice has attracted a lot of attention right after it was
proposed.
However in the last few
years it has become clear that before lattice methods could offer a concrete
answer on $f$ and on the validity of the static approximation
various systematic errors should first be understood. We
aspire to provide answers on some of these issues in this report. Specifically
we will be addressing ground state projection using smearing techniques,
volume and finite $a$ effects and extrapolation to the continuum limit.

\section{Smearing techniques}
The first issue
concerning the pseudoscalar correlator is whether the ground state can be
unambiguously obtained using smearing techniques \cite{step,fb1}.
A systematic dependence of $f$ and the
pseudoscalar mass $\tilde{M}_P$ on the wave function size was reported in ref.
\cite{Jap} where the effect was found to be more pronounced at larger
$\beta$'s. Here we use $\tilde{M}_P$ to denote the pseudoscalar mass
obtained in the static approximation since this is not the real mass
but the mass with the heavy quark mass subtracted.
To investigate this issue we performed a careful analysis at $\beta=6.26$ on a
$18^3 \times 48$
lattice with 43 configurations using six wave functions of
r.m.s. radius varying from $3.9a$ to $6.5a$. Three of these
(denoted by $G1, G2, G3$) are constructed
as linear combinations of a local  and a gaussian interpolating function,
whereas the remaining three are purely gaussian characterised by the two
parameters
$n$ and $\alpha$. These wave functions are constructed in a gauge invariant
manner and require a very small investment in CPU time.
The light quark mass was fixed at about twice the strange
quark mass.
In order to obtain $f$ we need a good signal for both the ``smeared -
smeared" (quark fields smeared both at source and sink), and the
``local - smeared" (quarks smeared at either source or sink) correlators.
Ground state projection can be detected as a plateau in the `local mass'
$\mu(t)=log(C(t)/C(t-a))$ of the correlator $C(t)$. For the ``smeared-smeared"
correlator, $C^{SS}(t)$, the plateaus are relatively easy to identify,
with
the best wave function resulting in a plateau as early as $t=2a$.
The masses obtained by fitting $C^{SS}(t)$ in the plateau yield
consistent values for $a\tilde{M}_P=0.565(8)$. On the other hand
identifying the plateaus for the ``local-smeared" correlator $C^{LS}(t)$
is much trickier as it can be seen from fig~1. The wave function $G1$
with the smallest r.m.s. radius reaches the plateau at a large value of $t$
where noise dominates. As
we increase the radius the plateaus set in at smaller $t$
until we reach an optimum radius giving rise to the best plateau. Increasing
the radius even further results
 again in reaching the asymptotic value at larger $t$ e.g. wave function with
$n=160,~\alpha=5$.

\begin{table}[hbt]
\caption[]{Lattices used in this work. N gives the number of independent
configurations. The inverse lattice spacing
$a^{-1}$ is obtained using  the $\rho$ meson mass ($a^{-1}_{M_{\rho}}$)
and, in square brackets, the string tension $\sqrt{\sigma}$ \cite{BaSchi}.
 The spatial extension
$L$ of the lattice was estimated using $a_{M_\rho}$.
}
\label{tab:lattices}
\begin{tabular}{lrrrrr}
\hline
 $\beta$ &  $L/a$ & $ L_t/a$ & N & $a^{-1}$/GeV & $L/fm$   \\
\hline
 &  4 &24 & 404 & & 0.55 \\
&  6 &24 & 131 &  & 0.82 \\
5.74 &  8&24 & 270 &1.45(19) &  1.10\\
&  10 &24 & 213 & [1.12] &  1.37\\
&  12 &24 & 140 &  &  1.65\\ \hline

5.82 &  6 &28 & 100 & 1.72\cite{fb1} & 0.70 \\ \hline

&  6 &36 & 227 &   & 0.53 \\
&  8 &36 & 100 &2.25(10)  &  0.70\\
6.00 & 12 &36 & 304 &[1.88]  &  1.05\\
& 18 &36 & 27 &  &  1.58\\ \hline

& 12 &48 & 103 &  3.70(32) & 0.64 \\
6.26 & 18 &48 & 76 &  &  0.96\\
\hline
\end{tabular}
\end{table}

Therefore without prior knowledge of the height of
the plateau one is liable to misidentify the position of the plateau and thus
arrive at wave function dependent results.
For this purpose, we have inserted into fig.~1 the plateau as obtained
from a fit to $C^{SS}(t)$
using the best  wave function with $\alpha=4,~n=100$.
As can be seen the t-dependent masses
finally end up  in the same plateau. The procedure of first fitting
the ``smeared-smeared" correlator to obtain the mass $\tilde{M}_P$ and then
using this value of $\tilde{M}_P$ in a constrained fit of the ``local-smeared"
correlator in the range of the plateau
yields the same result for $f$ (within error bars) for {\it all} wave
functions.

Having thus established the
wave function independence of the static results, for the rest of the lattices
we performed  the calculation using the best wave function.

\begin{figure}[htb]
\vspace{55mm}
\caption[]{Local mass for ``local - smeared" correlator for six different
wave functions at $\beta=6.26$.
The error band arises from
fitting $C^{SS}(t)$ for the best gaussian wave function $n=100,~\alpha=4$.}
\label{fig:largenenough}
\end{figure}

\section{The pseudoscalar decay constant}
Our systematic study of volume and finite $a$ effects
was performed using the lattices listed in
table~1. The quantity

\begin{equation}
\fhat = f \sqrt{M_P}
\left ( \frac{\alpha_s(e^{-2/3}M_P)}{\alpha_s(e^{-2/3}M_B)} \right ) ^{6/25}
\end{equation}
is considered since it has a finite limit as $M_P \rightarrow \infty$.
In order to check scaling we form the dimensionless ratio $\fhat/\sigma^{3/4}$
where the string tension  $\sigma$ is obtained from ref. \cite{BaSchi}.
We show our results at different $\beta$ values as a function of
$L\sqrt{\sigma}$ in fig.~2 with the light quark mass fixed so that the
corresponding pseudoscalar mass $M_P(l,l)$ satisfies
$M_P^2(l,l)/\sigma=4$.
If finite $a$ effects are negligible then all data should follow
a universal curve.
Since all data are statistically
independent
the relatively reasonable fit of the
data (dashed line in fig.~2) to a function depending solely on $L$ is
tantamount to $a$ effects smaller than the 10\% level
of the general error bars. For our most accurate data with errors smaller
than 10\%
finite $a$ effects are of course visible. From fig.~2 we also learn that finite
volume
effects are negligible beyond lengths of $L\sqrt{\sigma} \sim 3$ for the light
quark mass considered here. Using our accurate results at $\beta=5.74$
for the lattices with $L/a=12$ and $L/a=8$ we do not see any volume dependence
also for the lighter quark masses that we used in this study. We thus
can use our data at $L\sqrt{\sigma} \sim 3$ for which we have most statistics
for
a more detailed analysis.

\begin{figure}[htb]
\vspace{53mm}
\caption[]{$\hat{F}/\sigma^{3/4}$ vs $L\sqrt{\sigma}$ for the lattices of
table~1.
The light quark mass is fixed to about twice the strange quark mass.}
\label{fig:largenenough}
\end{figure}

We extrapolate to $\kappa_{critical}$ by fitting linearly our results for
$M_{\pi}^2$.
To fix the physical scale we use both the $\rho$ mass $M_{\rho}$
and the string tension $\sigma$. From our data we seem to find that using
$f_{\pi}$ is equivalent to using $M_{\rho}$ since $f_{\pi}/M_{\rho}$ is
within our accuracy independent of the light quark masses and consistent with
the
experimental value.
The $\kappa$ corresponding to
the strange quark mass is found by matching to the kaon mass.
In fig.~3 we show the results at $\beta=5.74,~6$ and 6.26 for the decay
constant at the chiral limit
using both $\sigma$ and $M_{\rho}$ to fix the scale. Equivalent plots can be
made for the strange quark mass. Setting the scale using $\sqrt{\sigma}$ makes
finite $a$ effects rather small and a linear extrapolation to $a=0$ gives
results within error bars of  the value obtained at $\beta=6.26$. On the
other hand when $M_{\rho}$ is used the errors are much larger. The final
(extrapolated) results are, however, in agreement for the two scales. We
obtain in $GeV^{3/2}$
\begin{equation}
\fhat_{s}= 0.61(4)(7), ~~ \fhat_{u}=0.53(5)(6)
\end{equation}
with $\sqrt{\sigma}$ and
\begin{equation}
\fhat_{s}= 0.67(18)(7), ~~ \fhat_{u}=0.59(18)(6)
\end{equation}
from the $M_{\rho}$ scale. The first errors are statistical and the second
result from the uncertainty in $Z_{stat}=0.71(8)$ which was determined by using
renormalised perturbation theory \cite{LeMa}.
At the B meson mass we obtain in MeV
\begin{equation}
f_{bs}= 266(22)(26), ~~ f_{bu}=230(22)(26)
\end{equation}
and at the D meson mass the extremely high value of $f_{cs}=405(27)(46)$~MeV
making $1/M_P$ corrections necessary.

\begin{figure*}[htb]
\vspace{50mm}
\caption[]{ $F=\fhat/Z_{stat}$ is shown vs $a$, in (a) in units of
$\sqrt{\sigma}$
and in (b) in units of $M_{\rho}$.}
\label{fig:largenenough}
\end{figure*}

\section{Mass splittings}
The mass difference between the scalar and the pseudoscalar,
$\Delta_S$, and between the $\Lambda$ baryon and the pseudoscalar,
$\Delta_{\Lambda}$, were determined directly by looking at the
ratio of the respective correlators.
The signal for $\Delta_S$ is reasonable
but not as good as for the pseudoscalar.
Therefore our results for $\Delta_S$ are not accurate enough
to allow a definite statement on the light quark mass dependence of this
quantity. The linear extrapolation to the continuum using
$\sqrt{\sigma}$ gives 344(37) MeV at quark mass, $m_{2s}$, twice that of the
strange
quark. For $\Delta_{\Lambda}$ we can only quote a rough estimate of 600 MeV
since our signals were not good enough for any extrapolation.

\section{String breaking}

Here we estimate the distance $R_b$ where the {\it full} QCD potential flattens
off using only quantities calculated in the quenched approximation
ref. \cite{Tsu,fb3}.
For distances up to about 0.7 fm the QCD potential is approximated rather well
by the quenched $Q\bar{Q}$ potential. Also at very large distances we expect
the potential to be well represented by the quenched approximation. The
asymptotic behaviour of the potential is given by the mass of two mesons in
the static approximation.
Thus we can obtain an upper bound for $R_b$
by setting $V(R_b)=2\tilde{M}_P$,
with $V(R)$ being the quenched potential \cite{BaSchi}
including the
self-energy term which cancels in the given relation.
In fig.~4 we
show the a-dependence of $R_b$ at the chiral limit measured in units of
$\sqrt{\sigma}$.
The corresponding figures for quark masses up to $m_{2s}$ show very similar
results.
Our extrapolation to $a=0$
gives for all quark masses  up to $m_{2s}$
\begin{equation}
 R_b=1.9(2)(2)~fm .
\end{equation}

\begin{figure}[htb]
\vspace{55mm}
\caption[]{ $R_b/\sqrt{\sigma}$ vs $a\sqrt{\sigma}$ at the chiral limit}
\label{fig:largenenough}
\end{figure}

Because of the weak dependence of this quantity on the light quark mass, we
expect the screening of the potential in full QCD to appear at distances
that are rather independent of the dynamical quark mass and hence can be
studied with relatively large quark masses.

\end{document}